\documentclass[twocolumn,showpacs,floatfix,prl]{revtex4}

\usepackage{graphicx}
\usepackage{dcolumn}
\usepackage{bm}
\usepackage{amsmath}
\usepackage{amssymb}

\begin{document}

\title{Commensurability effects for fermionic atoms trapped in 1D optical lattices}

\author{Rafael A. Molina$^1$}
\author{Jorge Dukelsky$^1$}
\author{Peter Schmitteckert$^2$}

\affiliation{$^1$Instituto de Estructura de la Materia - CSIC, Serrano 123,
28006 Madrid, Spain \\
$^2$Institut f\"ur Theorie der Kondensierten Materie,
Universit\"at Karlsruhe, 76128 Karlsruhe, Germany}

\begin{abstract}
Fermionic atoms in two different hyperfine states confined in optical lattices show strong commensurability
effects due to the interplay between the atomic density wave (ADW) ordering and the lattice potential.
We show that spatially separated regions of commensurable and
incommensurable phases can coexist. The commensurability between the harmonic trap and the lattice sites can be used
to control the amplitude of the atomic density waves in the central region of the trap.
\end{abstract}


\maketitle

Ultracold Bose or Fermi gases can be confined in artificial optical
lattices created by standing-wave laser fields.
The low-energy properties of these systems can be described using
models borrowed from condensed matter systems\cite{Cirac03} whose
parameters and dimensionality can be controlled with high precision.
This tunability has opened new avenues for understanding
the physics of strongly correlated systems. Greiner {\em et al.} observed a
superfluid to Mott insulator
transition in a 3D optical lattice with bosonic $^{87}Rb$ atoms.\cite{Greiner02}
Interesting experimental results have also been obtained
for fermions.\cite{Stoferle06}

The physics of cold fermionic atoms in optical lattices is predicted to be the
one of the attractive Hubbard model with the different hyperfine states
playing the role of the spin states. In the
case of one dimension, bosonization predicts the formation of a Luther-Emery
liquid for attractive interactions.\cite{LutherEmery} The spin sector is
gapped, inducing exponential decay of spin correlations in contrast to
singlet superconducting and charge-density wave correlations that have a power
law decay.\cite{Giamarchi} Atoms are trapped in experiments and the interplay
between correlations, lattice, and confinement has to be properly
addressed. Moreover, the study of the inhomogeneity can be very relevant for
the understanding of unconventional
superconductors. For example, the density of states of high-T$_c$
superconductors presents spatial inhomogeneities.\cite{Hanaguri04}
Confined fermions in optical lattices are ideal experimental candidates to
study these issues.

Recently, Gao Xianlong {\em et al.} studied the unpolarized
attractive Hubbard model in a 1D optical lattice with harmonic confinement.
Using the density matrix renormalization group (DMRG) approach,
they observed the coexistence of spin pairing with an ADW
that could be interpreted as a signature of a Luther-Emery liquid phase.
This interpretation is further justified by the fact that the momentum of
the ADW is proportional to the average
atomic density in the bulk of the trap $k_{ADW}= \pi \bar{n}$.
These ADWs could be detected measuring the elastic light-scattering diffraction
pattern (the Fraunhofer structure
factor), proportional to the Fourier transform of the atomic
density.\cite{Gao07} In a complementary work
using Quantum Monte Carlo simulations, Karim Pour {\it et al.} interpreted
the divergence of the form factors of
the density-density and pairing correlation function as defining a
supersolid.\cite{Karim}

The purpose of this Letter is to uncover new commensurability effects
induced by the combination of the lattice and confinement
potentials and the atomic density. Since the
precise momentum of the ADWs in the trap can be controlled with the average
density, it is possible to observe effects similar to the
commensurability-incommensurability transition appearing in crystalline
surfaces when the density oscillations have a different periodicity than
the underlying lattice.\cite{Pokrovsky79} These concepts have also been
applied to the doped Mott transition in strongly correlated Fermi systems.\cite{Giamarchi91,Giamarchi}
We will show that for local
densities close to half-filling there appear commensurate and incommensurate
phases in different parts of the
lattice. Similar local quantum criticality issues have been seen before in
the repulsive case.\cite{Rigol03}
In the commensurate phase that arises in sectors of the lattice where the density
is close to half-filling, the
amplitude of the ADWs is enhanced due to {\em local umklapp} scattering terms.
The incommensurate phase close to
half-filling is characterized by the appearance of a new length scale,
similar to a beating length, due to the interplay between the
periodicity of the lattice and the periodicity of the ADW
that results in nodes and amplitude modulation of the latter.

Similar to previous works, we consider the Hamiltonian
\begin{eqnarray}
\label{eq:H}
\hat{H}= & -t\sum_{i,\sigma}
\left(\hat{c}^{\phantom{\dagger}}_{i\sigma}\hat{c}_{i+1\sigma}+h.c.\right)
+U\sum_i\hat{n}_{i\uparrow}\hat{n}_{i\downarrow} \nonumber \\
& +V\sum_i\left(i-L/2+D\right)^2\hat{n}_i ~, \label{Hub}
\end{eqnarray}
where $t$ is the hopping, $\sigma$ is a
pseudospin-1/2 degree of freedom, $U$ is the interaction
(always attractive $U<0$), and $V$ is the strength of the confinement.
The creation, destruction and number operators are the usual ones
at each site $i$ of the lattice.
The total length $L$ of the system is chosen such that the density is smooth,
going to zero in the edges. All
energies are expressed in units of $t$ ($t=1$). The parameter $D$ measures
the displacement of the center of the harmonic confinement potential with
respect to the position of a lattice site. It can vary from $0.0$ (the center
of the harmonic well coincides with a lattice site) to $0.5$ (the center of
the well is exactly between two lattice sites). We will study only the
unpolarized case.

We study the ground state properties of the Hamiltonian (\ref{Hub}) with the DMRG algorithm\cite{DMRG} that
provides very accurate numerical results. Due to the breaking of the translational symmetry by the trapping
potential some modifications of the DMRG procedure are required. Here we follow the same procedure used before for
disorder potentials\cite{schmitteckert} and umklapp scattering induced phase
transitions.\cite{schmitteckert_werner} In order to obtain enough accuracy for the largest size systems we needed
to keep a maximum of 1400 states in each iteration.

As a complementary tool we use the approximate Hartree-Fock (HF) method. It will allow us to study system sizes
significantly beyond the limits of a DMRG calculation.  The HF approximation describes with high quantitative
precision the ADWs for $|U| < 1$ as we have checked for systems up to $L=250$ by comparison with DMRG results. For
values of $U$ between $U=-1$ and $U=-2$ the HF description is qualitatively good for the ADWs but it overestimates
their amplitude. For $|U| > 2$ HF deviates from the exact results due to strong pairing correlations not included
in the approximation.

\begin{figure}
\begin{center}
\includegraphics[width=8cm]{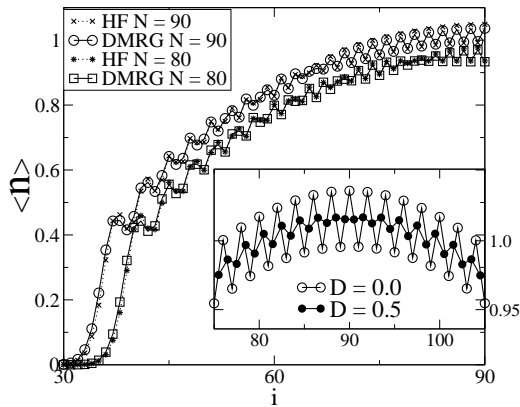}
\caption{\label{fig:statwave2} DMRG and HF results for $L=180$, $V=0.0005$, $D=0.0$, $U=-1.0$, and $N=80, 90$. As
the graph is symmetric only the left part is shown. The inset shows a zoom of the central region for $N=90$
comparing DMRG results with $D=0.0$ and $D=0.5$. }
\end{center}
\end{figure}

Without the trapping potential bosonization predicts an ADW with wavenumber
$k_\text{ADW}=2k_\text{F}$\cite{Giamarchi}. Adding confinement the wave number of the ADW is modulated by the
density profile in the trap, but it can still be written as a function of an effective
Fermi wavenumber $k_\text{F}^\text{eff}$%
\cite{Gao07}
\begin{equation}
\label{eq:kADW} k_\text{ADW}=2k_\text{F}^\text{eff}, ~\ k_\text{F}^\text{eff}=\pi \bar{n}/2,
\end{equation}
where $\bar{n}$ is the average density around the center of the trap. For $\bar{n}=1.0$ (half-filling)
$k_\text{ADW}=\pi$ and we have an ADW with a periodicity of two lattice sites.

Figure ~\ref{fig:statwave2} shows DMRG results for the site density of a system with $L=180$, $U=-1$, $V=0.0005$,
$D=0.0$ and two values of the number of atoms $N=N_{\uparrow}+N_{\downarrow}$. $N=80$ is represented by open
squares and $N=90$ by open circles. We also show for comparison HF results for the same system. The agreement
between DMRG and HF is remarkable for $U=-1$. As can be seen in the figure, the local density is close to
half-filling in the center of the trap. As a consequence, an ADW with a long range modulation of the amplitudes
develops. The effect is more pronounced in the case with $N=90$, displaying a nice commensurate ADW with large
amplitude. The density in the center of the trap is below half-filling for $N=80$, however the ADW can still be
seen.

\begin{figure}[t]
\includegraphics[width=6.5cm]{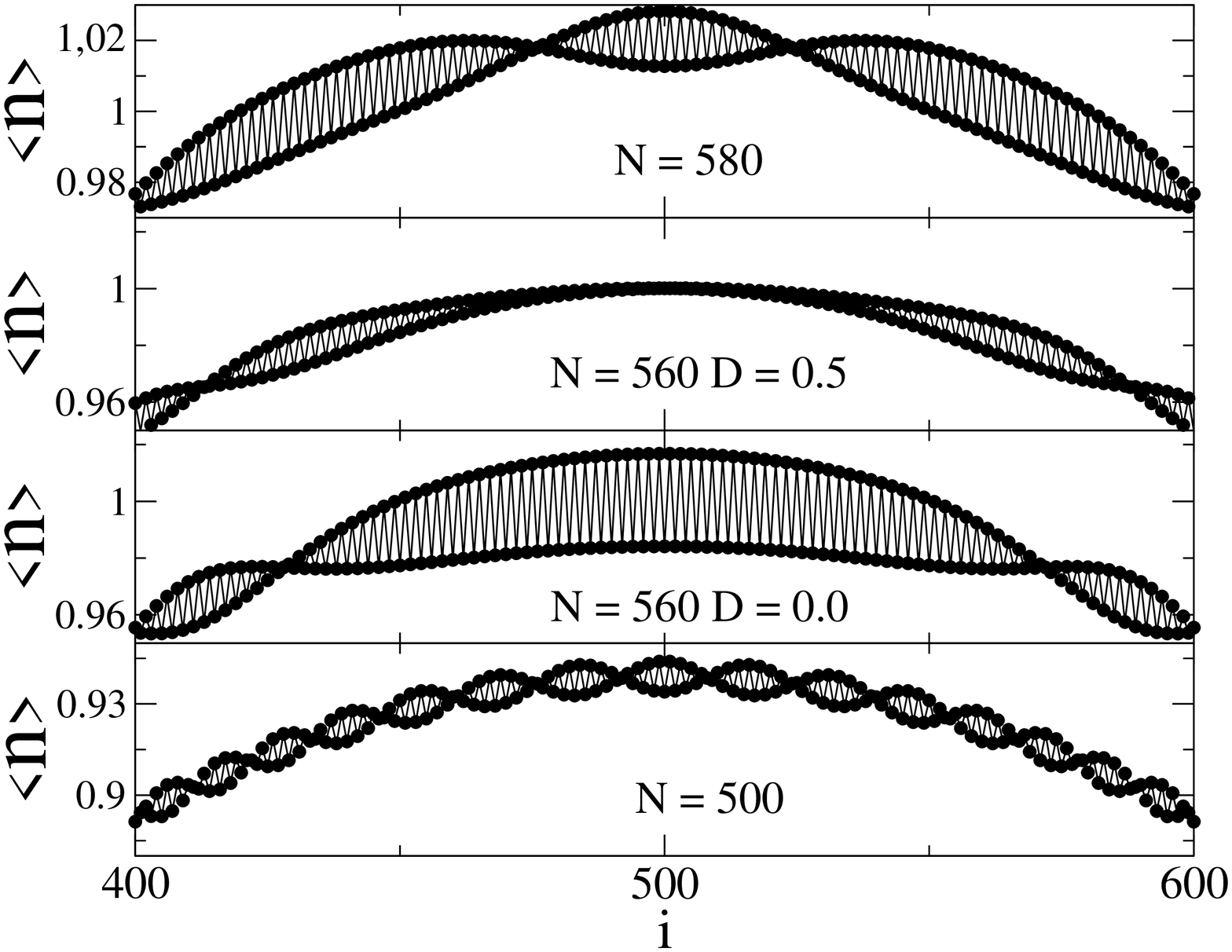}
\caption{\label{fig:500} HF local density as a function of the site index for a system of $L=1000$ sites and
 different values of $N$ and $D$.}
\end{figure}

The appearance of lobes and nodes in the ADW depends on small deviations of $k_{ADW}$ from the lattice momentum
$\pi$. These features are expected to be more pronounced for larger lattices. Therefore, we resort to the HF
approximation to treat lattice sizes that are not attainable with present DMRG codes. In order to relate
intermediate and large lattice systems we use the scaling $NV^2=$ constant while $N \rightarrow \infty$ and $V
\rightarrow 0$. \cite{Damle96} This scaling keeps constant the density in the central region of the trap. In
Fig.~\ref{fig:500} we show HF results for the density in the central region for systems with $L=1000$,
$V=0.000012$, $U=-1$ and different number of atoms, to illustrate the kind of ADWs appearing in larger size
systems as a function of the number of atoms. Of particular interest is the case of $N=560$ where we can see the
suppression of the ADW for a displacement $D=0.5$ of the trap.

The distance between nodes $\delta$ in the ADW is related to the distance it takes for the system to realize it is
not commensurate with the underlying lattice. If the local average density is close to half-filling $\bar{n}=1.
\pm \epsilon$, we have from (\ref{eq:kADW}) $k_{ADW}=\pi (1 \pm \epsilon)$. $\delta$ can be from the distance
between nodes for a wave $\cos[(\pi(1 \pm \epsilon)x]$ taken at integer values of the variable $x$, leading to
$\delta=\epsilon^{-1}$.

\begin{figure}
\begin{center}
\includegraphics[width=7cm]{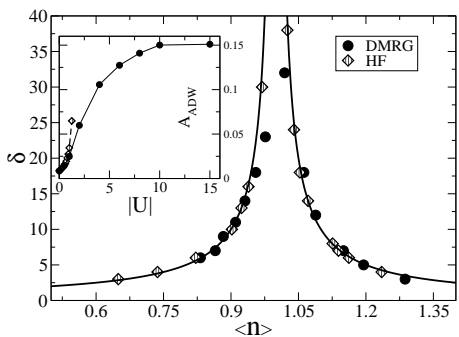}
\caption{\label{fig:statwave2b}  Distance between nodes ($\delta$) as a function of the average value of the
density in the center of the trap for HF calculations with $L=1000$, $V=0.000012$, and $U=-1.0$, and DMRG results
with $L=100$, $V=0.001$, and $U=-4.0$. Inset: Amplitude of the ADW $A_{ADW}$ in the center of the trap for
different values of the interaction strength with the central density at half-filling, $L=100$, and $V=0.001$.
DMRG results (black dots), HF results for interaction strenghts up to $U=-1.5$ (diamonds). }
\end{center}
\end{figure}

Figure ~\ref{fig:statwave2b} shows HF and DMRG results of the distance between nodes $\delta$ vs. the average
local density in the center of the trap. The continuous line represents the formula $\delta=\epsilon^{-1}$. Due to
 finite size effects, slight deviations could be seen close to half-filling. However, the overall agreement is excellent.

 The enhancement of the ADW amplitude close to half-filling can be explained in terms of the additional umklapp
interaction. In the bosonization language these effects are treated by adding a term to the umklapp part of the
Hamiltonian.\cite{Giamarchi}
\begin{equation}
\label{eq:umklapp}
H_{\delta}=g_3\int dx \cos\left(\sqrt{8}\phi(x)-\delta x\right),
\end{equation}
where $g_3$ is the coupling constant of the umklapp scattering terms, proportional to $U$ in our model, and
$\phi(x)$ is the bosonized field. The additional lowering of the energy due to the umklapp term is effective even
for local densities not precisely at half-filling. This phenomenon is robust close to half-filling, giving rise to
the enhancement of the  amplitude and range modulation of the ADW. The amplitude of the ADW increases with the
value of the attractive interaction as shown in the inset of Fig.~\ref{fig:statwave2b} for DMRG and HF
calculations with $L=100$ and $V=0.001$ but changing the number of particles for each value of the interaction to
keep local half-filling in the trap center. The increase is very abrupt for small values of $|U|$, saturating for
$|U| \sim 10.$

Another important source of commensurability comes from the external trapping potential. Just at half-filling a
phase difference in the ADW can have dramatic effects in the observed amplitudes at the lattices sites. The
central part of the ADW for $N=90$ is shown in the inset of Fig.~\ref{fig:statwave2}, comparing the results with
$D=0.0$ and $D=0.5$. The strong suppression of the ADW can be clearly seen. Even though the properties of the
Luther-Emery liquid are not modified by the displacement of the external potential, the actual position of the
lattice sites determine the amplitudes of the ADW. This property was not seen previously in the
literature\cite{Machida06}, because the systems studied were too small to notice the difference but can have big
experimental consequences for the observation of the ADWs. Figure~\ref{fig:500} also shows these differences
between $D=0.0$ and $D=0.5$ for a larger system.

\begin{figure}
\includegraphics[width=7cm]{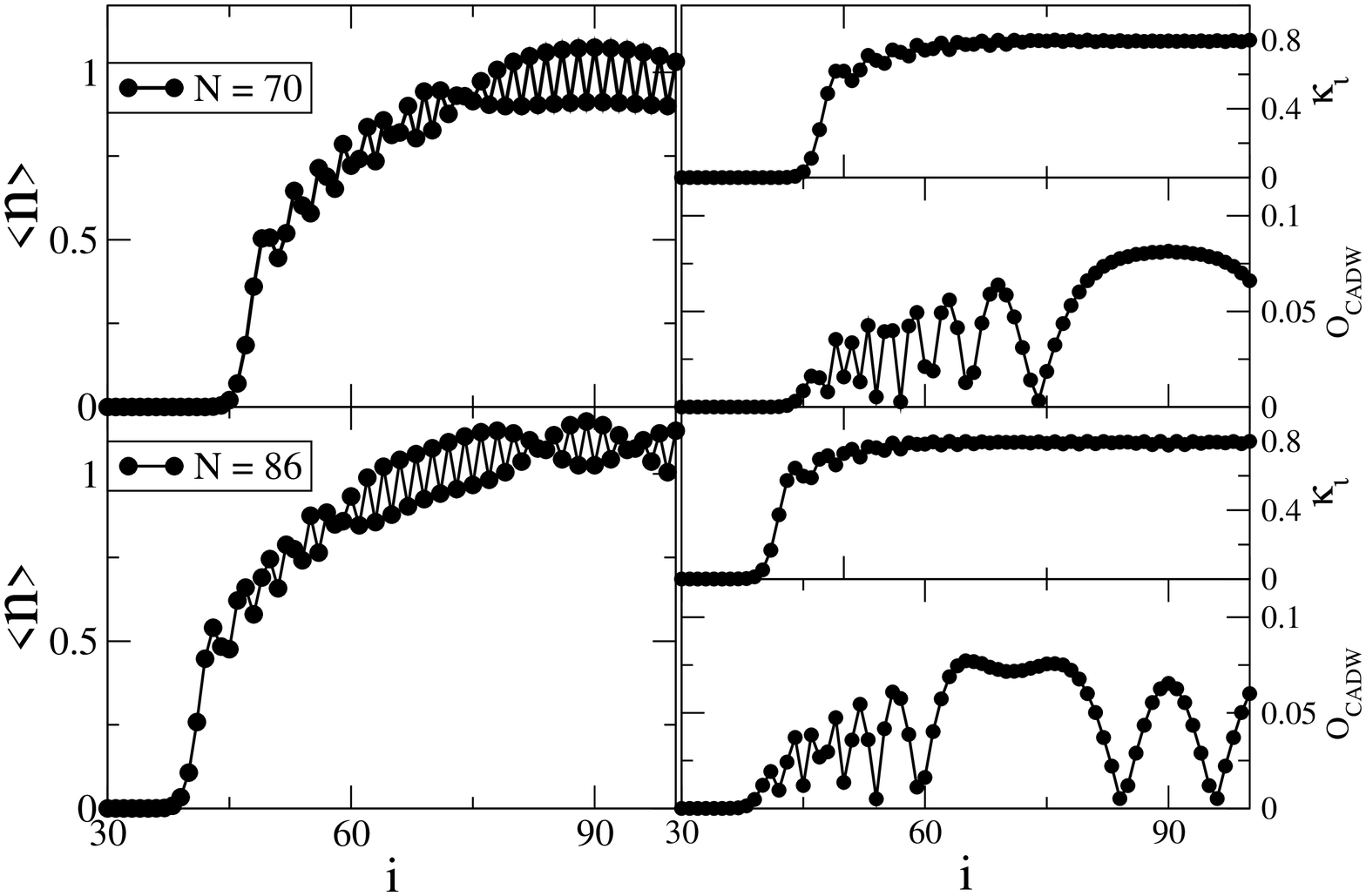}
\caption{\label{fig:zero} DMRG results for $L=180$, $D=0.0$, $V=0.0005$, and $U=-4$, {\em Top panel:} $86$, atoms.
{\em Bottom panel:} $N=70$. The left panels show the density and the right panels show the local order parameters
explained in the text.}
\end{figure}

To proceed further with the characterization of the different local phases
and commensurability effects we will
make use of two local quantities, the local variance of the density
$\kappa_i=\left<n_i^2\right>- \left<n_i\right>^2$ and the order parameter
$O_{CADW}(i)$
for a local  commensurate (pinned) ADW  defined as:
\begin{equation}
\label{eq:OCDAW}
O_{CADW}(i)=\left|\left<n_i-\frac{1}{2}(n_{i+1}+n_{i-1})\right>\right| ~.
\end{equation}
While the local variance $\kappa_i$ measures the fraction of spin paired particles in site $i$,  the commensurate
ADW order parameter $O_{CADW}(i)$ measures the local amplitude of the ADW between site $i$ and its near neighbors.

In Fig.~\ref{fig:zero} we show DMRG results for $L=180$, $V=0.0005$, $U=-4$, $N=70$, and $N=86$. The amplitude of
the ADW is much higher than in the examples with $U=-1$. The density in the case with $N=70$ is very close to
half-filling in the center and the amplitude of the ADW very large there. The local umklapp scattering term
effectively increases the attraction between fermions close to half-filling. The other case has densities larger
than $n=1.0$ in the central region. Nodes in the density due to the incommensurate ADWs appear. The amplitude of
the ADWs is larger in the lobe with local density close to half-filling and so it is the distance between nodes
there. In the right panels we show the value of $\kappa_i$ and $O_{CADW}(i)$ for both examples. The results are
close to a plateau in the central region with a large $\kappa_i=0.8$. A more careful examination reveals that the
maximum of $\kappa_i$ occurs in the regions close to half-filling, precisely where $O_{CADW}(i)$ has a maximum.

\begin{figure}
\begin{center}
\includegraphics[width=7cm,height=5cm]{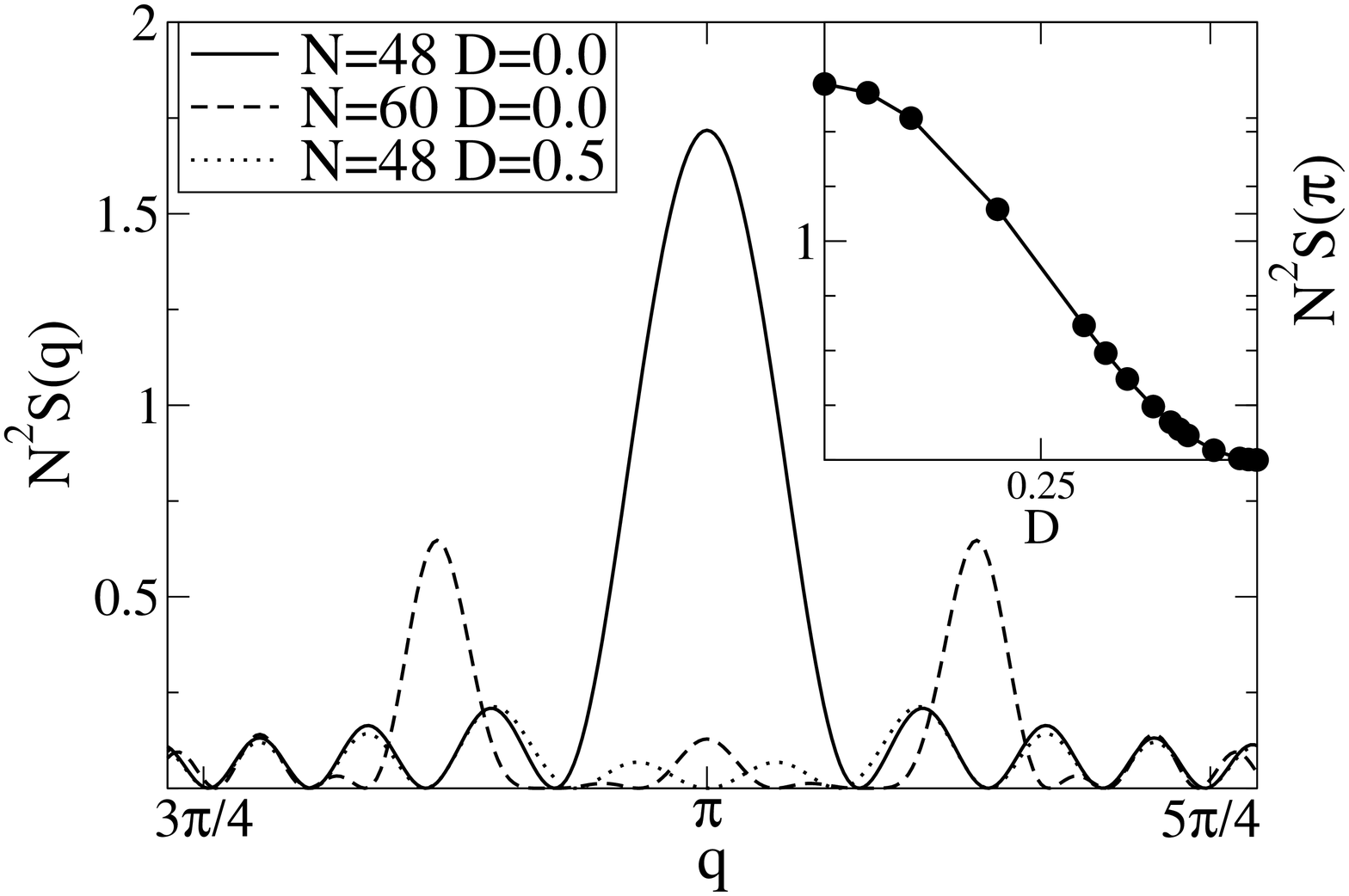}
\caption{\label{fig:sq} Structure factor $S(q)$ for different DMRG examples
with $L=100$ and $V=0.001$. In the inset we show the value of the height
of the peak at $q=\pi$ as a function of $D$ for the example with $N=48$.}
\end{center}
\end{figure}

Lobes and nodes in the ADWs can be observed through elastic light-scattering diffraction experiments, by means of
the Fraunhofer structure factor\cite{LutherEmery}:
\begin{equation}
S(q)=\frac{1}{N^2} \left|\sum_j \exp{(-iqj)}n_j\right|^2.
\end{equation}
Fig.~\ref{fig:sq} shows the structure factor for $L=100$, $V=0.001$, and $U=-4$. We show three cases with $N=60$
$D=0.0$, $N=48$ $D=0.0$, and $N=48$ $D=0.5$. largest peak at $q=\pi$ corresponds to the case $N=48$ and $D=0.0$,
exactly at half-filling. This is due to the large amplitude commensurate ADW. For the same number of particles and
$D=0.5$, the ADW disappears and $S(q)$ is suppressed in the region around $q=\pi$. Above half-filling ($N=60$) a
small peak shows up at $q=\pi$, while most of the intensity concentrates at the peaks $q_{max}=\pi\bar{n}$ and the
corresponding symmetric peak. In the inset we show the value of $S(\pi)$ as a function of $D$ to demonstrate that
the displacement of the trap is an observable effect.

In conclusion we have shown that commensurate and incommensurate phases can be seen in cold fermions trapped in
optical lattices with attractive interaction. The two phases can appear spatially separated depending on the
density and confinement of the atoms in the lattice. The amplitude of the ADW increases with the strength of the
interaction. The local density variance $\kappa_i$ shows that the commensurate phase is characterized by a maximum
number of local spin-paired atoms. The structure factor studied in different examples reveals that the
commensurate phase is characterized by a large peak exactly at $k=\pi$, while for phase coexistence the intensity
is distributed in several peaks. The displacement of the trap in relation to the lattice is also a very important
parameter for the experiment. Depending on it, the amplitude of the ADWs and the $q=\pi$ peak can be suppressed.
The width of the peaks is inversely proportional to the number of wells in the optical lattice. This parameter is
the main limitation for the experimental detection of this commensurability effects. To resolve the two peaks
appearing due to the incommensurability of ADWs with densities $\bar{n}=1.0 \pm\epsilon$ one needs widths of the
peaks smaller than $\Gamma < \pi \epsilon$. According to our numerical simulations, for harmonic confinement, the
intrinsic width of the peak scales with size as $\Gamma_{int} \sim 10/L$. Taking into account that the width
induced by the experimental set-up $\Gamma_{exp}$ will convolute with the intrinsic width we shall need a system
size of $L \gtrsim 10/\sqrt{4\epsilon^2/\pi^2-\Gamma_{exp}^2}$ to resolve the two peaks. Currently available
one-dimensional optical lattices of 100 sites should be enough to detect the results represented in
Fig.~\ref{fig:sq} if the experimental width is less than 0.24 in units of the density\cite{Stoferle06}.

The authors acknowledge discussions with M. Rodriguez,
D. Weinmann, R.M. Noack, S. Rombouts, A. Rela\~no, and
J. Gonz\'alez Carmona. This work
is supported in part by Spanish Government grant
Ref. FIS2006-12783-C03-01 and by grant CAM-CSIC Ref. 200650M012.
RAM contract is financed by CSIC and
the European Comission through the I3p program.

\end{document}